\documentclass[12pt]{article}

\usepackage{newtxtext,newtxmath}
\usepackage{graphicx}

\usepackage[letterpaper,margin=1in]{geometry}

\renewenvironment{abstract}
	{\quotation}
	{\endquotation}

\date{}

\makeatletter
\renewcommand{\fnum@figure}{\textbf{Figure \thefigure}}
\renewcommand{\fnum@table}{\textbf{Table \thetable}}
\makeatother

\usepackage{scicite}

\usepackage{url}

\def\scititle{
	Seeing Forests Through Clouds
}
\title{\bfseries \boldmath \scititle}

\author{
	Anastassia M. Makarieva$^{1,2\ast}$,
	Andrei V. Nefiodov$^{1,2}$, \and
	Antonio D. Nobre$^{3}$, \and
	Luz A. Cuartas$^{4}$,\and
	Paulo Nobre$^{3}$,\and
	Germ\'{a}n Poveda$^{5}$, \and
	Jos\'{e} A. Marengo$^{4}$, \and
	Anja Rammig$^{6}$, \and
	Susan A. Masino$^{7}$, \and
	Ugo Bardi$^{8}$, \and
	Juan F. Salazar$^{9}$, \and	
	William R. Moomaw$^{10}$, \and
	Scott R. Saleska$^{11}$\and
	\small$^{1}$Theoretical Physics Division, Petersburg Nuclear Physics Institute of NRC {\textquotedblleft}Kurchatov Institute{\textquotedblright}, \\ \small 188300 Gatchina, Russia. \and
	\small$^{2}$Institute for Advanced Study, Technical University of Munich, 85748 Garching, Germany.\and
		\small$^{3}$National Institute for Space Research (INPE), S\~{a}o Jos\'{e} dos Campos, S\~{a}o Paulo 12227-010, Brazil. \and
		\small$^{4}$National Center for Monitoring and Early Warning of Natural Disasters (CEMADEN), S\~{a}o Jos\'{e} dos Campos, \\ \small S\~{a}o Paulo 12227-010, Brazil. \and
		\small$^{5}$Department of Geosciences and Environment, Universidad Nacional de Colombia, Medell{\'\i}n, Colombia. \and	
		\small$^{6}$School of Life Sciences, Technical University of Munich, 85354 Freising, Germany.\and
		\small$^{7}$Neuroscience Program and Psychology Department, Life Sciences Center, Trinity College, Hartford, CT 06106, USA.\and
		\small$^{8}$Department of Chemistry, University of Florence, Firenze, Italy.\and
		\small$^{9}$Escuela Ambiental, Facultad de Ingenier{\'\i}a, Universidad de Antioquia, Medell{\'\i}n, Colombia.\and		
		\small$^{10}$Global Development and Environment Institute, Tufts University, Medford, MA 02144, USA. \and	
		\small$^{11}$Department of Ecology and Evolutionary Biology, University of Arizona, Tucson, AZ 85719, USA.\and
	\small$^\ast$Corresponding author. Email: ammakarieva@gmail.com\and
}

\begin{document} 

% Insert the title and author list
\maketitle

% Abstract, in bold
% There are strict length limits, and not all formats have abstracts.
% Consult the journal instructions to authors for details.
% Do not cite any references in the abstract.
\begin{abstract} \bfseries \boldmath
Goessling \textit{et al.} \cite{goessling2024} link the record-breaking warming anomaly of 2023 to a global albedo decline due to reduced low-level cloud cover. What caused the reduction remains unclear. Goessling \textit{et al.} considered several geophysical mechanisms, including ocean surface warming and declining aerosol emissions, but did not discuss the biosphere. We propose that disruption of global biospheric functioning could be a cause, as supported by three lines of evidence that have not yet been jointly considered.
\end{abstract}

% The first paragraph of any Science paper does NOT have a heading
% Nor is it indented
\noindent
First, plant functioning plays a key role in cloud formation \cite{zhao2017,dror-schwartz2021,cerasoli2021,duveiller2021,xu2022,ellison2024}. In one model study, converting land from swamp to desert raised global temperature by $8$~K due to reduced cloud cover \cite{lague2023}. In the Amazon, the low-level cloud cover increases markedly with the photosynthetic activity of the underlying forest \cite{heiblum2014}.

Second, in 2023, photosynthesis on land experienced a globally significant disruption, as signalled by the complete disappearance of the terrestrial carbon sink \cite{ke2024}. Terrestrial ecosystems, which typically absorb approximately one-fourth of anthropogenic CO$_2$ emissions, anomalously ceased this function. This breakdown was attributed to Canadian wildfires and the record-breaking drought in the Amazon \cite{espinoza2024}.

Third, Goessling \textit{et al.}  focus on changes over oceans, but their maps show that some of the largest reductions in cloud cover in 2023 were over land, including over Amazonian and Congolian forests. Another cloud reduction hotspot is evident over Canada. Besides, precipitation over land in 2023 had a major negative anomaly, $-0.08$~mm\,day$^{-1}$ \cite{adler2024}.

Growing pressure on forests is known to induce nonlinear feedbacks, including abrupt changes in ecosystem functioning \cite{zemp17b,makarieva23,flores2024}. Feedbacks of similar strength in global climate models are unknown \cite{boos2016}. The biospheric breakdown in 2023 may have triggered massive cloud cover reduction facilitating the abrupt warming. 

If verified, the good news is that the recent extra warmth could wane if the forests partially self-recover. With the many unknowns remaining, we urge more integrative thinking and emphasize the importance of urgently curbing forest exploitation to stabilize both the climate and the biosphere \cite{moomaw2019,makarieva23b}.

%%%%%%%%%%%%%%%% REFERENCES %%%%%%%%%%%%%%%

\clearpage % Clear all remaining figures and tables then start a new page

% The list of references goes after the main text and before the acknowledgements
% When preparing an initial submission, we recommend you use BibTeX, like this:
%
%\bibliography{sci_forest} % for a file named science_template.bib

\begin{thebibliography}{10}
\providecommand{\url}[1]{\texttt{#1}}
\expandafter\ifx\csname urlstyle\endcsname\relax
  \providecommand{\doi}[1]{doi:\discretionary{}{}{}#1}\else
  \providecommand{\doi}{doi:\discretionary{}{}{}\begingroup
  \urlstyle{rm}\Url}\fi

\bibitem{goessling2024}
H.~F. Goessling, T.~Rackow, T.~Jung, Recent global temperature surge
  intensified by record-low planetary albedo. \emph{Science}
  \textbf{387}~(6729), 68--73 (2024), \doi{10.1126/science.adq7280}.

\bibitem{zhao2017}
D.~F. Zhao, \emph{et~al.}, Environmental conditions regulate the impact of
  plants on cloud formation. \emph{Nat. Commun.} \textbf{8}~(1), 14067 (2017),
  \doi{10.1038/ncomms14067}.

\bibitem{dror-schwartz2021}
T.~Dror-Schwartz, I.~Koren, O.~Altaratz, R.~Heiblum, On the abundance and
  common properties of continental, organized shallow (green) clouds.
  \emph{IEEE Trans. Geosci. Remote Sens.} \textbf{59}~(6), 4570--4578 (2021),
  \doi{10.1109/TGRS.2020.3023085}.

\bibitem{cerasoli2021}
S.~Cerasoli, J.~Yin, A.~Porporato, Cloud cooling effects of afforestation and
  reforestation at midlatitudes. \emph{Proc. Natl. Acad. Sci. U.S.A.}
  \textbf{118}~(33), e2026241118 (2021), \doi{10.1073/pnas.2026241118}.

\bibitem{duveiller2021}
G.~Duveiller, \emph{et~al.}, Revealing the widespread potential of forests to
  increase low level cloud cover. \emph{Nat. Commun.} \textbf{12}, 4337 (2021),
  \doi{10.1038/s41467-021-24551-5}.

\bibitem{xu2022}
R.~Xu, \emph{et~al.}, Contrasting impacts of forests on cloud cover based on
  satellite observations. \emph{Nat. Commun.} \textbf{13}, 670 (2022),
  \doi{10.1038/s41467-022-28161-7}.

\bibitem{ellison2024}
D.~Ellison, J.~Pokorn{\'y}, M.~Wild, Even cooler insights: {On} the power of
  forests to (water the {Earth} and) cool the planet. \emph{Glob. Change Biol.}
  \textbf{30}~(2), e17195 (2024), \doi{10.1111/gcb.17195}.

\bibitem{lague2023}
M.~M. Lagu\"{e}, G.~R. Quetin, W.~R. Boos, Reduced terrestrial evaporation
  increases atmospheric water vapor by generating cloud feedbacks.
  \emph{Environ. Res. Lett.} \textbf{18}~(7), 074021 (2023),
  \doi{10.1088/1748-9326/acdbe1}.

\bibitem{heiblum2014}
R.~H. Heiblum, I.~Koren, G.~Feingold, On the link between {Amazonian} forest
  properties and shallow cumulus cloud fields. \emph{Atmos. Chem. Phys.}
  \textbf{14}~(12), 6063--6074 (2014), \doi{10.5194/acp-14-6063-2014}.

\bibitem{ke2024}
P.~Ke, \emph{et~al.}, Low latency carbon budget analysis reveals a large
  decline of the land carbon sink in 2023. \emph{Natl. Sci. Rev.}
  \textbf{11}~(12), nwae367 (2024), \doi{10.1093/nsr/nwae367}.

\bibitem{espinoza2024}
J.-C. Espinoza, \emph{et~al.}, The new record of drought and warmth in the
  {Amazon} in 2023 related to regional and global climatic features. \emph{Sci.
  Rep.} \textbf{14}~(1), 8107 (2024), \doi{10.1038/s41598-024-58782-5}.

\bibitem{adler2024}
R.~F. Adler, G.~Gu, Global precipitation for the year 2023 and how it relates
  to longer term variations and trends. \emph{Atmosphere} \textbf{15}~(5), 535
  (2024), \doi{10.3390/atmos15050535}.

\bibitem{zemp17b}
D.~C. Zemp, \emph{et~al.}, Self-amplified {Amazon} forest loss due to
  vegetation-atmosphere feedbacks. \emph{Nat. Commun.} \textbf{8}, 14681
  (2017), \doi{10.1038/ncomms14681}.

\bibitem{makarieva23}
A.~M. Makarieva, \emph{et~al.}, The role of ecosystem transpiration in creating
  alternate moisture regimes by influencing atmospheric moisture convergence.
  \emph{Glob. Change Biol.} \textbf{29}~(9), 2536--2556 (2023),
  \doi{10.1111/gcb.16644}.

\bibitem{flores2024}
B.~M. Flores, \emph{et~al.}, Critical transitions in the {Amazon} forest
  system. \emph{Nature} \textbf{626}~(7999), 555--564 (2024),
  \doi{10.1038/s41586-023-06970-0}.

\bibitem{boos2016}
W.~R. Boos, T.~Storelvmo, Reply to {Levermann et al.: Linear} scaling for
  monsoons based on well-verified balance between adiabatic cooling and latent
  heat release. \emph{Proc. Natl. Acad. Sci. U.S.A.} \textbf{113}~(17),
  E2350--E2351 (2016), \doi{10.1073/pnas.1603626113}.

\bibitem{moomaw2019}
W.~R. Moomaw, S.~A. Masino, E.~K. Faison, Intact forests in the {United States:
  Proforestation} mitigates climate change and serves the greatest good.
  \emph{Front. For. Glob. Change} \textbf{2} (2019),
  \doi{10.3389/ffgc.2019.00027}.

\bibitem{makarieva23b}
A.~M. Makarieva, A.~V. Nefiodov, A.~Rammig, A.~D. Nobre, Re-appraisal of the
  global climatic role of natural forests for improved climate projections and
  policies. \emph{Front. For. Glob. Change} \textbf{6} (2023),
  \doi{10.3389/ffgc.2023.1150191}.

\end{thebibliography}
%\bibliographystyle{sciencemag}

% After the paper has completed peer review and been revised ready for acceptance,
% you should comment out the lines above and copy-paste the contents of your .bbl
% file here instead. This will help ensure that our conversion software works correctly.
% Remember to re-run BibTeX first - check the timestamp!
%
% Example of the first three entries copy-pasted from science_template.bbl:
%
%\begin{thebibliography}{1}
%
%\bibitem{example}
%A.~N. {Author}, An example reference. \emph{Journal of Improbable Research}
%  \textbf{1}, 67 (2020).
%
%\bibitem{example2}
%F.~M. {Surname}, S.~{Author}, A second example. \emph{Interesting Research
%  Letters} \textbf{32}, 897 (2019).
%
%\bibitem{example_preprint}
%P.~{One}, P.~{Two}, P.~{Three}, {An unpublished preprint}. \emph{preprint}
%  (2021), arXiv:2101.12345.
%
%\end{thebibliography}

%%%%%%%%%%%%%%%% ACKNOWLEDGEMENTS %%%%%%%%%%%%%%%

%\section*{Acknowledgments}

\paragraph*{Funding:}
A.M.M. is partially funded by the Federal Ministry of Education and Research (BMBF) and the Free State of Bavaria under the Excellence Strategy of the Federal Government and the L\"ander, as well as by the Technical University of Munich -- Institute for Advanced Study.

%%%%%%%%%%%%%%%% END OF MAIN TEXT %%%%%%%%%%%%%%%

\end{document}